# Development of NavIC synchronized fully automated inter-building QKD framework and demonstration of quantum secured video calling

Adarsh Jain[1], Abhishek Khanna, Jay Bhatt, Parthkumar V Sakhiya, Shashank Kumar, Rohan S Urdhwareshe, Nilesh M Desai

*Space Applications Center (SAC), Indian Space Research Organization (ISRO),*
*Jodhpur Tekra, Ambawadi Vistar P.O. Ahmedabad – 380015, Gujarat, India*

## Abstract

Quantum key distribution (QKD) is a revolutionary communication technology that promises ultimate security assurance by exploiting the fundamental principles of quantum mechanics. In this work, we report design and development of a fully automated inter-building QKD framework for generation and distribution of cryptographic keys, securely and seamlessly, by executing weak coherent pulse based BB84 protocol. This framework is experimentally validated by establishing a quantum communication link between two buildings separated by ~300m of free-space atmospheric channel. A novel synchronization technique enabled with indigenous NavIC (IRNSS) constellation is developed and implemented. This QKD system demonstrates generation of secure key rate as high as 300 Kbps with QBER< 3% for mean photon no. per pulse (μ) of 0.15. The intercept-resend eavesdropping attack has been emulated within the system and evaluated during experiment. A novel quantum secured end-to-end encrypted video calling app (QuViC) is also developed and integrated with QKD framework to demonstrate unconditionally secure two-way communication over Ethernet, functioning alongside with quantum communication. [1]

## 1 Introduction

Secure communication and data security perceived as essential commodity in recent times with deployment in wide area of applications. In a booming world of digitization, an unconditional security is the need of the hour and can be achieved by quantum key distribution (QKD), proposed as a method to generate unconditionally secure shared cryptographic keys for two distant parties. It is one of the most mature applications of quantum communication, which in contrast to conventional cryptographic technologies is not dependent on mathematical and computational complexities. The secure and symmetrically grown key is used to encrypt message signal enabling user to communicate with absolute security. The presence of an Eavesdropper over quantum channel introduces detectable errors, which reflects in the quantum bit error rate (QBER).

QKD has come a long way since its proposal in 1984 [1] and first table-top experiment in 1989[2]. So far, many researchers have demonstrated various QKD protocols based on fiber and free space QKD. But fiber based and terrestrial free space QKD are limited to short distance[3, 4], due to absorption in fiber cables or atmospheric losses. The satellite based free space QKD link[5, 6] can overcome such limitations by successfully distributing secure keys between two ground stations situated over thousands of kilometers apart. QKD technology has become mature over the period of time, however detailed & efficient software framework for QKD is not discussed much. This work tries to bridge this gap by providing details of a fully automated end-to-end QKD framework capable of generation and distribution of encryption keys between two remote buildings followed by secure communication over classical channel. Earlier, to demonstrate secure communication, quantum secured video call has been done with AES encryption between two different ground station[7] and one-time-padding (OTP) encryption-based quantum video surveillance[8], only after the key accumulation. Our QKD architecture integrates video calling app with software itself, whereby along with key generation it can also be used for secure video calling by employing quantum cryptography. This paper presents the design and development of fully automated NavIC synchronized inter-building QKD framework on LabVIEW platform. Earlier authors reported[9] the demonstration of QKD framework within close lab environment by physical mode of synchronization. However, this QKD framework is capable of automated handshaking, qubit exchange, time and frame synchronization using NavIC receivers based on IRNSS[10] along with classical communication. The performance of the QKD framework is validated by demonstrating BB84 protocol over a distance of 300m of atmospheric channel and achieved ∼300 Kbps of secure key rate for a QBER <3 %. The synchronizations between Alice and Bob is achieved by implementing a novel synchronization technique using NavIC constellation. The software is also equipped with an emulator which mimics an eavesdropping attack. This QKD system is also shown to have the capability of distributing secure keys over a distance of upto few kilometers with key rates of the order of tens of Kbps. For real time application, the framework is integrated with a novel video calling app (QuViC), encrypted by quantum keys, can function at the same time along with QKD.

---
[1]adarshjain@sac.isro.gov.in

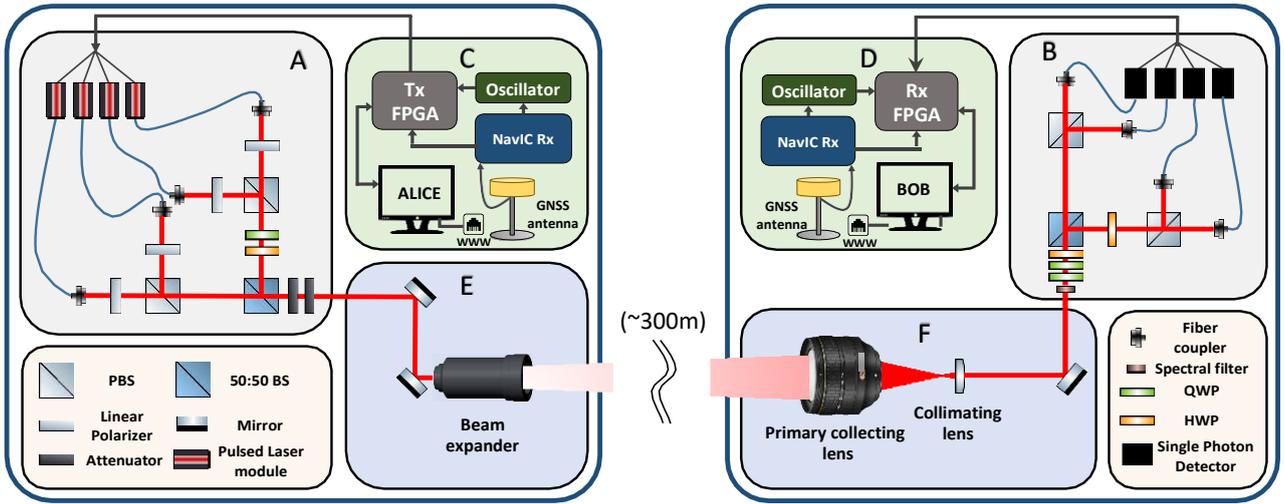

Figure 1: Inter-building QKD framework with free space quantum communication link over 300m (A) Quantum Tx module consists of pulsed laser modules (PLM) along with BB84 encoding optics module based on weak coherent pulses (B) Quantum Rx module consists of BB84 decoding optics module along with single photon detectors (SPD) (C) &(D) These sections involve FPGA modules connected to PLMs and SPDs . NavIC receiver based subsystems for synchronization, PC connected to Ethernet (E) Front end optics of the transmitter terminal and (F) receiver terminal for efficient beam transmission and reception.

This paper is organized as follows. In section II, system design methodology is presented in detail. Section III describes NavIC enabled synchronizations and section IV describes software implementation strategy for QKD framework. Section V describes inter-building free space QKD link experiment followed by results and discussion in section VI. Section VII discusses quantum secured end-to-end video calling application, QuViC. Finally, we conclude and outline the future work in section VIII.

## 2 System Design

In this section, the outlay of the overall inter-building QKD system based on BB84[1] protocol along with key rate estimation is presented.

### 2.1 QKD framework

The inter-building QKD framework for free space quantum communication over 300 m of atmospheric channel is depicted in Fig 1. The pulsed laser modules (PLMs) along with BB84 encoding optics module transmits a stream of polarization encoded photons. Neutral density (Nd) filters were used to set the overall attenuation of source so as to keep output mean photon number (MPN) close to 0.15 to suppress the multi photon events. The beam is then directed towards the front end optics module for transmission into the free space. The Beam expander along with one adjacent mirror is mounted on a gimbal system for controlling the azimuth and elevation orientation of the transmitted beam. The PLMs are randomly triggered with FPGA connected to Alice's PC.

The quantum Rx consists of front end optics module followed by BB84 decoding optics module and single photon detectors (SPD). The front end optics provides a collimated beam to BB84 decoding optics module. The polarized photons are then detected using SPDs. The optics assembly shown in B & F section is mounted on a two-axis rotary mount. The receiver module is compactly enclosed in a blackout enclosure to suppress the dark noise due to stray light.

The visible beacon source, with 638nm wavelength is used for coarse alignment between Alice and Bob terminals. Then the fine alignment is achieved by appropriate movements of Tx gimbal & Rx rotary mount. The NavIC enabled synchronization mechanism is used for synchronizing Alice and Bob terminals. The aerial view of the two buildings inside SAC, ISRO campus for free space QKD experiment is shown in Fig 2. The Alice and Bob QKD terminals were deployed atop these buildings.

### 2.2 Key Rate Estimation

For Bob terminal, the transmission efficiency of front-end optics and polarization decoding module was measured as 66.7% and 75.47% respectively. The average detection efficiency of SPD is taken as 70%, as per the test data. Overall pointing and atmospheric loss was measured to be less than 0.5dB, bring the overall calculated efficiency of the quantum channel to ∼31.64%.

For a secure communication the MPN should be of the order of efficiency of channel $O(\eta)$. To be on safer side, the MPN chosen was ∼0.15. For weak coherent pulse (WCP) source of a MPN $\mu$ and total efficiency $\eta$, the key can be estimated as follows:

A WCP source follows poissonian statistics, probability of i-photon state is:

$$P_i = \frac{\mu^i}{i!}e^{-\mu} \quad (1)$$

Probability $\eta_i$, that pulse containing i photons will trigger the detector, assuming all photons are independent of each other's behaviour in i-photon state is given by

$$\eta_i = 1 - (1-\eta)^i \quad (2)$$

Hence probability ($Q_i$) that there is i-photon state and it triggers the detector is given by

$$Q_i = \eta_i \times P_i \quad (3)$$

Then the overall probability that a WCP source of MPN= $\mu$ will trigger a detector is [11]

$$Q_\mu = \Sigma_{i=0}^{\infty} \eta_i \times P_i = 1 - e^{-\eta\mu} \quad (4)$$

Hence if the pulse repetition frequency (PRF) of laser is $f$, then expected clicks per second is given by

$$E_{clicks} = f \times Q_\mu \quad (5)$$

which results in $E_{clicks} \approx$ 920K for PRF=20MHz.

Before post processing, 0.5% i.e. 100k elements out of total 20M sampled data elements are utilized by the Rx side front-end software for overall frame synchronization bringing the synchronization efficiency to 99.5% such that the total number of clicks available for sifting and key distillation is $E_{clicks}$ (after synchronization) ∼915K. Further post processing which includes key sifting, QBER calculation, error correction and privacy amplification [12, 13] as an overall efficiency of 32.65%. Hence the expected secure key rate is 299 Kbps as tabulated in table 1 .

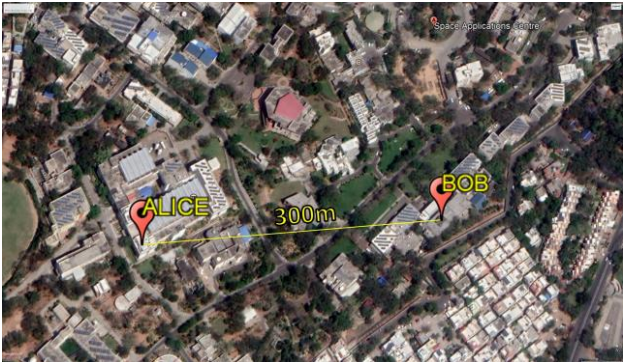

Figure 2: Aerial view of the two buildings inside SAC campus for free space QKD experiment.

Table 1: Key rate estimation

| Sr. | Parameters | Value |
|---|---|---|
| 1. | MPN | 0.15 |
| 2. | PRF | 20MHz |
| 3. | Channel Efficiency | 31.64% |
| 4. | Synchronization efficiency | 99.5% |
| 5. | Post processing efficiency | 32.65% |
| 6. | Total clicks | ∼ 920K |
| 7. | Click (after sync.) | ∼915K |
| 8. | Secure key rate | ∼299Kbps |

## 3 NavIC enabled synchronization technique

As the quantum Tx and Rx are separated by large distance and have independent reference clocks, overall timing synchronization is critical before labelling QKD photon pulse sequence for the key sifting and correlation process. For this purpose, a synchronized time base is required at both the communicating ends. In our framework, the source is preparing quantum states at rates of 20 MHz corresponding to 50ns timebase, while the detection rates can be much lower because of low mean photon number and path attenuation. The objective of the synchronization procedures is to identify which photon detection event at the receiver corresponds to which photon emitted from the source and have an automated framework continuously generating quantum secured keys without errors.

NavIC is an Indian regional navigation satellite constellation system developed and operated by Indian Space Research Organisation ISRO[10]. Each NavIC satellite has an atomic clock and all atomic clocks in NavIC satellites are synchronized periodically by the ground control segment, which monitors clock errors and updates them to maintain the accuracy of the NavIC system. NavIC receivers lock and process the signals from these satellites and estimate position and time information. After the estimation, the receiver clocks are precisely synchronized with the NavIC system time and generate a 1PPS (pulse-per-second) signal along with the corresponding system time information. These PPS pulses can be used as absolute time marks for timing purposes and to discipline the reference clock sources at different locations for clock time-base synchronization.

A chip-scale atomic clock (CSAC) is used for the reference 10 MHz clock (oscillator), which offers an excellent short-term stability with Allan deviation of 3e-10 at 1 sec. This clock is disciplined using NavIC 1PPS signal for long-term operations and clock synchronization.

As Alice and Bob terminals are stationary, the time-of-flight for the photons is fixed. There are following two primary challenges involved in QKD timing synchronization:

1. Clock synchronization (time-base synchronization)

2. Frame synchronization (absolute time tagging)

For the purpose of clock synchronization, reference clock of FPGAs at both the ends is disciplined using NavIC Rx [14]. The reference clock is multiplied inside Bob's FPGA to 80 MHz. This gives a resolution of 12.5 ns in time-tagging at photon detection events. For frame synchronization, the start pulse for photon generation at $T_x$ and counting at Rx is synchronized using 1PPS pulses, called as absolute time tagging as the primary start of the counters is synchronized. There are primarily two sources of errors between Alice and Bob based on these NavIC based synchronization methods:

## 3.1 Relative clock jitter after synchro-nization

The instantaneous change of frequency between two ends causes some of the photons to be tagged across the time-bins, each having resolution of 12.5 ns. It was observed that, with synchronized clocks the spread of the photon detection events can be between 3 bins ($\pm$1 bin from the central bin after correcting time of transmission) which corresponds to 37.5 ns. Hence, to mitigate this issue of the photon detection time spread, nearest neighbor correlation (NNC) algorithm has been deployed inside the FPGA i.e. back-end during data acquisition process, presented subsequently in section IV.

## 3.2 1PPS timing jitter

Practically 1-PPS signal obtained from two NavIC receivers can be relatively inaccurate having typical jitter of $\sim$50ns, which may occasionally increase up to $\sim$100 ns. This can lead to incorrect detection event tagging due of frame overlap and frame hopping issues, presented in section IV, after each new run. Hence, to mitigate these inaccuracies, we performed coincidence matching consisting of time-tagging and synchronization [15] for which additional processing at the back-end as well as the front-end is required to achieve accurate frame synchronization.

# 4 Development of Software suite for QKD

The BB84 QKD protocol based end-to-end custom software suite has been developed, catering to the need of generating and distributing quantum keys between two remote parties in real-time, using NavIC based synchronization. Both the back-end and front-end software solution for overall synchronization, system control and monitoring has been implemented using LabVIEW platform.

For overall framework implementation, a state machine architecture has been followed and a GUI panel is developed for configuring the system settings of QKD framework as well as monitoring and logging of all the critical performance parameters in real-time. Besides this, as mentioned in section II, few additional back-end and front-end procedures required to enable software compensation of NavIC reference jitter and realize a robust field deployable synchronization mechanism have also been implemented. The software procedure can be described as follows:

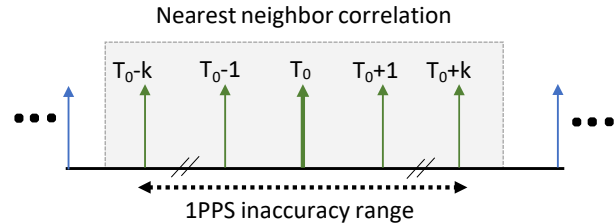

Figure 3: Nearest neighbor correlation algorithm implementation.

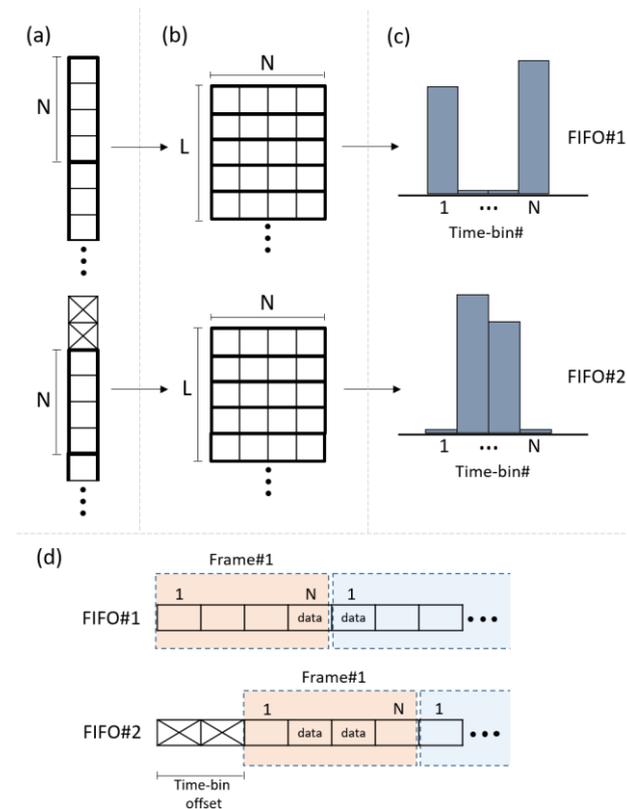

Figure 4: Illustration of auto-frame boundary selection for mitigation of frame overflow scenario: (a) Two separate raw data string obtained from Rx FIFO memory is (b) first divided into L (equal to pulse repetition rate) blocks, each having N (time-bins/frame) elements (c) Bin-wise histogram is obtained for both the datasets and (d) raw data from one of the two FIFO sets is selected.

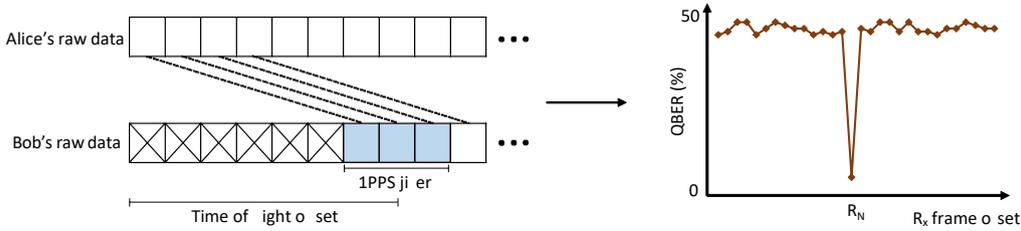

Figure 5: Illustration of Auto-frame offset selection procedure to uniquely determine the overall Rx-frame offset: $R_N$.

## 4.1 Qubit Exchange

**Transmission procedure:**

A pseudo random binary sequence of word length of 11-bits (PRBS11) is used for generation of two random bits each for Alice's bases and bits inside the Tx FPGA program. Based on the assigned duty cycle and PRF, one of the PLMs corresponding to a PRBS is then triggered.

**Detection procedure**

The Rx side FPGA, continuously samples all four channels connected to the SPDs, generating 18ns TTL output pulse per click event, at the rate of 80MHz.
For both the procedures, all the relevant parameters required for post-processing are transferred into corresponding FIFO memory.

**NNC based clock synchronization**

Besides, reference clocks at both ends being disciplined using 1PPS signal, NNC algorithm is implemented inside Rx-side FPGA during real-time data acquisition to mitigate any undesired photon detection time spread observed due to relative drift of reference clocks. . In this algorithm, every transmit side event is associated with corresponding receiver side photon detection event, registered at not just a single time-bin but also its nearest neighbor (adjacent) time-bins, within the estimated photon detection time spread, as shown in Fig. 3. With this method, key rates achievable is at par with having a physically common reference clock scenario. However, slight increase in QBER is observed due to more number of detection events containing dark counts as well as background counts now being considered within each frame. Once the reference clocks are synchronized, the acquired data is ready for further post-processing.

## 4.2 Post processing

Once qubit exchange is done and all the relevant data is transferred to corresponding FIFO memories, a predefined chunk of data from acquired dataset is read at the Alice ($T_x$) PC and Bob (Rx) PC terminals and bit unpacking is performed to get an array containing information about all the relevant link parameters like the individual channel status (Ch. 1-4), bases, bits, multi-photon/no-photon events registered (at Rx) etc. Following post-processing procedures are implemented over this data to get the final secure key.

**Frame synchronization:**

Firstly, frame offset at the Rx terminal due to time of flight delay needs to be compensated for. Besides that, due to the relative 1PPS timing inaccuracy of upto 100ns, the start pulse variation between two terminals can be such that the actual single photon detection event data may get registered simultaneously in adjacent frame also. This overflow across two adjacent frames is referred to as frame overflow condition. Moreover, given the timing inaccuracy of 1PPS can be larger than entire frame size itself, care needs to be taken for such frame hopping scenarios, where detection event of a given frame may get erroneously recorded in any of the adjacent frames. These frame synchronization requirements have been tackled using following back-end/front-end post processing techniques:

### 4.2.1 Auto frame boundary selection:

In this Rx terminal only technique, to mitigate frame overflow conditions, firstly the back-end code prepares two separate raw data strings: FIFO#1 and FIFO#2, consisting of recorded detection event per SPD and export it to Rx FIFO for further processing, as shown in Fig. 4. The FIFO#1 has no delay while FIFO#2 is prepared assuming delay of two time-bins i.e. 25ns, while registering click events. In either case, NNC algorithm is first implemented over the data before it is ready to be pushed into separate FIFO memories.
Thereafter, at the front-end, bin-wise time histogram of detection events is created to monitor any potential frame overflow condition and automatic selection of data from one of two FIFOs for further processing is performed.

### 4.2.2 Auto frame offset selection:

Beside the Rx frame offset expected due to time of flight delay, there can be max. 8 bins within the ambiguity interval of max. 100 ns of two NavIC receivers. Therefore, for tackling frame hopping condition, once the raw data from correct FIFO has been fetched, a small subset of the raw data is mutually shared between Alice and Bob. Thereafter, cross correlation of the Rx side data w.r.t $T_x$ side data is iteratively performed with incremental delays of 50ns (equivalent to 1 full frame at 20MHz PRF) and an estimation of interim QBER is carried out as shown Fig 5. Finally, a minimum-QBER search algorithm is employed to determine an appropriate Rx frame offset value, $R_N$, which allows to perform accurate frame-offset estimation for establishing raw key at the Bob side.

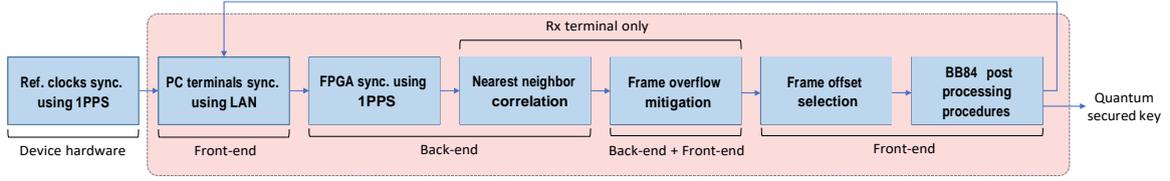

Figure 6: Flow diagram of overall synchronization mechanism between two remote terminals for realizing an automated QKD framework.

*Raw Key Generation:* Once, frame synchronization procedures are followed, the string containing the bits randomly selected at the Alice's setup and the deduced bits at the Bob's setup can be considered as the raw key.

*Sifted key generation:* Raw key preparation is followed by basis sharing by Bob over the public channel. Both the parties then discard all their bits where bases mismatch has occurred to form the sifted key. Once the key sifting procedure is completed, the QBER estimation is done using ∼5% of sifted key and presence/absence of Eve is verified.

*Quantum key generation:* The sifted key is used to perform key distillation procedures primarily consisting of error correction and privacy amplification[12, 13] to generate secure and error free quantum key. Error correction is done using implementation of Winnow protocol for 2-bit error correction. Error correction is followed by privacy amplification procedure based on the Toeplitz matrix, whereby every 16-bit block is compressed to just 11 bits, so that the final privacy amplified key is much shorter and considered as the final secure key or the quantum key.

Both the parties iteratively concatenate the current secure key with previously generated secure key until size of the final quantum key is at least equal to the desired size or size of the data itself that needs to be encrypted using the OTP technique. The overall synchronization mechanism for realizing an automated QKD framework is shown in Fig.6.

## 5 Experimental set-up for Inter-building QKD Framework

This inter-building QKD framework implements the BB84 protocol employing WCP source. Alice's terminal (Tx), as shown in Fig. 7a consists of quantum Tx, having indigenously developed pulse laser modules (PLM) [16] transmitting polarization encoded photons, with 20MHz pulse repetition rate and 5ns pulse width, at 785nm wavelength. These PLMs are randomly triggered using FPGA. This input beam passes through BB84 encoding optics module consisting of half wave plate (HWP) and polarization beam splitters (PBS), used to form two basis sets viz. rectilinear and diagonal. In order to maintain high polarization extinction ratio (PER), linear polarizers are used with each PLMs. The beams are co-aligned and combined using 50:50 beam splitter (BS). The beam is then directed to a beam expander using fold mirrors. This weak coherent beam is then passed through a window cut-out towards the Bob's terminal (Rx).

Bob terminal consists of front end optics followed by BB84 decoding optics module as shown Fig.7b. The collecting optics is a two lens telescopic setup, developed with a primary lens of 80mm aperture having focal length of 360mm and secondary lens of 25mm aperture and 25mm focal length. The Rx beam is having a 50mm footprint at collecting optics thus giving near 100% collecting efficiency. The observed beam divergence of the transmitted beam is $66\mu rad$. The incoming beam is collected and collimated to a beam cross section of∼ 4mm. The collimated beam first passes through the polarization compensation optics consisting of a combination of two quarter & one half waveplate. This is used to correct the additional phase delay introduced by the transmitting optics assembly as well as folding mirrors in Rx optics assembly. Once the polarization is corrected, the beam is passed through an optical filter of 5 nm FWHM bandwidth to cut down the ambient noise. The beam then passes through a 50:50 BS, which act as a random basis selector between rectilinear{ $|0^o\rangle$, $+90^o\rangle$ } and diagonal { $|-45^o\rangle$, $+45^o\rangle$ } basis sets. A HWP and PBS is used to form the polarization measurement basis. The transmitted path of the BS acts as a diagonally measurement basis whereas reflected path performs the measurement on rectilinear basis set. These polarizations decoded photons are then detected by SPDs and click events are recorded the Rx side FPGA.

Initially, the coarse alignment is done using a visible beacon laser of 638 nm wavelength at the Rx side and subsequently any Tx side 785nm laser with low attenuation is utilized for realizing fine alignment where gimbal platform at Alice's terminal and rotary stage at Bob's terminal help precisely controlling the orientation in azimuth and elevation direc-tions, thereby achieving LoS. Alice and Bob terminals were synchronized using NavIC receivers and also connected to Ethernet for carrying out classical communication over TCP/IP.

# 6 Results and Discussion

The experimental setup as described above is used to establish a free space quantum communication link between Alice and Bob terminals over 300m of atmospheric channel. QKD system employs linearly polarized photons to carry information and hence Alice and Bob must share common reference frames of polarization. However, due to multiple reflections and overall phase delay introduced by the front end optics assembly degrades polarization states originally prepared by the Tx module and introduce polarization errors. Any deviation from the linear polarization states increases errors in terms of QBER and must be corrected before measurement to guarantee a low-noise and reliable quantum communication. A combination of two quarter & one half wave plate is used to as polarization corrector [17] at the input of receiver. Fig. 8 shows the measured four polarization states before and after polarization correction and it can be observed that the received polarization states are satisfactorily restored back to originally prepared states.

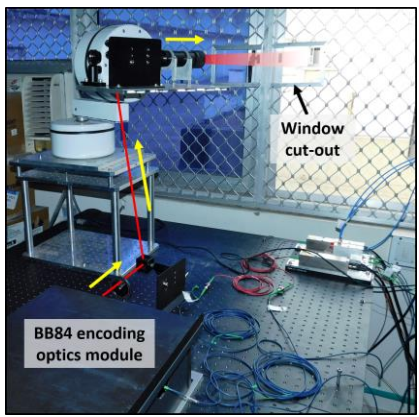

(a)

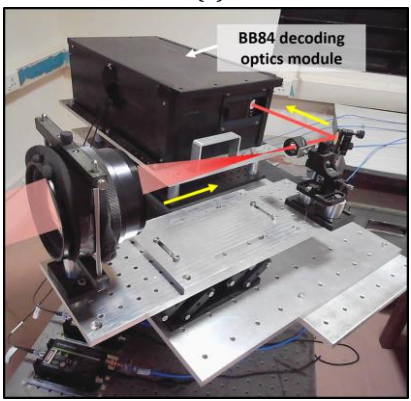

(b)

Figure 7: (a) Alice's terminal housed atop building A: the incoming beam from quantum encoding module is directed to beam expander, the beam propagates to free space via a cut out present in the window (b) Bob's terminal housed atop building B, collecting optics is a two lens setup. The beam is further folded and put into the quantum decoding module.

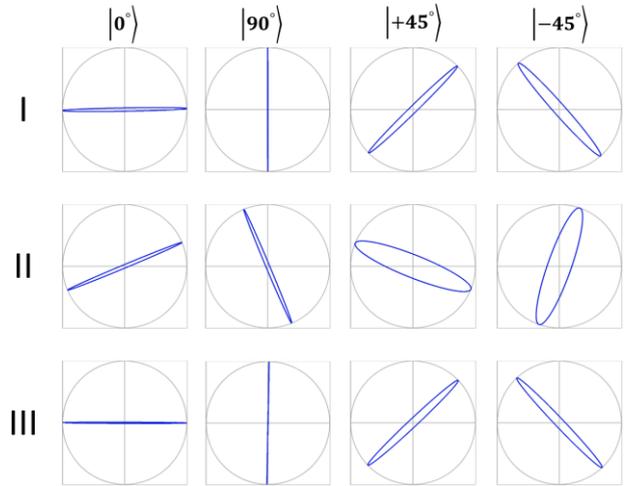

Figure 8: Measured polarization states (I) At output of polarization encoding module of Tx. (II & III) Before & after polarization correction at Rx

The experiment was carried out at night time. Before establishing quantum communication link, dark counts were measured and ensured <300 counts per second for quantum link to get operational.

The software code for BB84 QKD system on Tx and Rx side is optimized to operate at burst mode for quantum link and Public link. The 1-s bursts are dedicated for quantum link and timing synchronization (clock + frame synchronization) using NavIC receiver is performed for every burst. The interval between the successive bursts varies depending on processing capability and the available memory at Alice's and Bob's PC. Once clock and frame synchronization is achieved, actual BB84 QKD protocol state machine takes over. Regarding timing synchronization performance, Fig 9a shows that the frame overflow mitigation procedures helped reduce upto 45% discard events while Fig 9b shows the performance of autoframe offset correction procedure.

Furthermore, it can be deduced from the Rx-frame offset estimation that the average time of flight delay between two terminals corresponding to 20 frames of Rx side FPGA was appx. 1000ns, which then expected, translates to 300m free space distance between two building. Thus, this synchronization mechanism has proven its efficacy for realizing an automated QKD framework especially when working with practical devices having some inherent jitter.

The quantum communication link was established multiple times during nights for several hours and performance was found stable. For each experimental run, we record the sifted key generation rate, estimated QBER and final secure key rate as shown in Fig 10a. The measured QBER was stable and found to be ∼2.6%, with the sifted key rate ∼420 Kbps as shown in Fig 10a. The final avg. secure key rate after performing error correction and privacy amplification procedures is measured ∼280 Kbps, which is close to our keyrate estimation as discussed in section ??. This nonzero QBER in absence of eavesdropping can be attributed to several factors such as the dark counts, ambient noise, imperfect optical components etc. The slight variation in the sifted key rate and hence, in quantum key rate can be attributed to overall building vibrations, short term stability of gimbal at Tx end and rotary stage at Rx end could have further added to key rate variations.

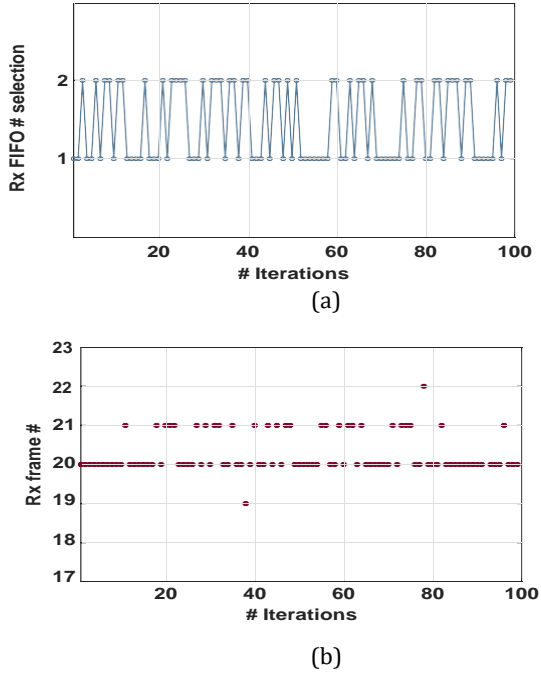

Figure 9: Frame synchronization procedures (a) Auto-frame overflow mitigation (b) Rx frame offset identification using auto-frame offset correction to account for 1PPS jitter and time of flight.

Experimental runs were also carried out with intercept-and-resend type of eavesdropping attack. The software code for realistic emulation of this type of attack is implemented inside Tx FPGA. When Eve is emulated to be present in the quantum channel, the QBER got increased to ∼25%, as shown in Fig. 10b. and sign of abort pops on GUI.

The key rate analysis of our QKD framework with existing Alice and Bob terminal is also carried out to evaluate its performance over varying transmission distances, as shown in Fig.11. It can be seen that even though due to the line of sight limitations between two buildings we could test it only upto 300m, our QKD framework is capable of generating secure keys at an average rate of ∼290 Kbps over a distance of upto 750m. Also, this analysis shows that the key rates of the order of about 50 Kbps are achievable for a distance of upto ∼2.5Km assuming negligible atmospheric absorption and similar pointing accuracy.

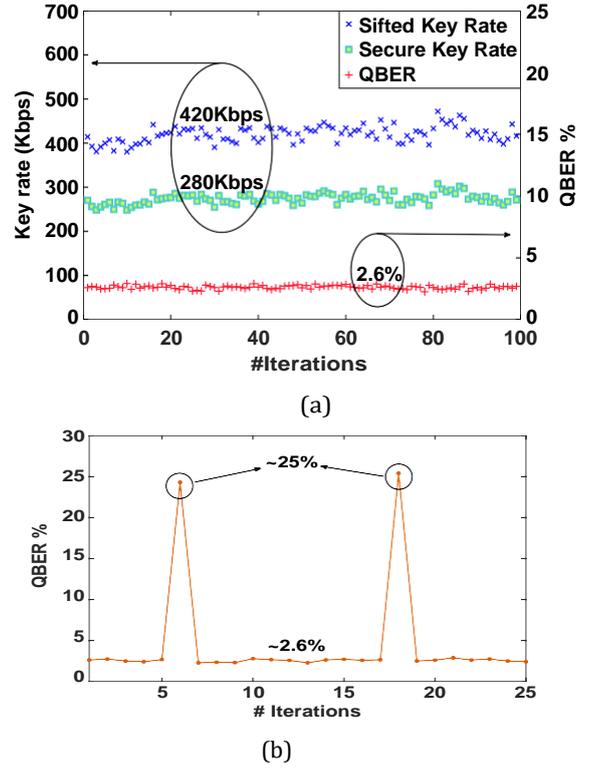

Figure 10: (a)Achieved sifted and secure key rates and QBER performance of inter-building QKD link. (b) Experimentally estimated QBER performance with emulated eavesdropping attack, stable QBER shoots up to ∼25% in the presence of Eve.

Therefore, this inter-building QKD framework, with Alice and Bob terminals having a very compact front-end optics, can be field deployed and sufficiently high key rates can be achieved catering to the requirements of practical cryptographic applications.

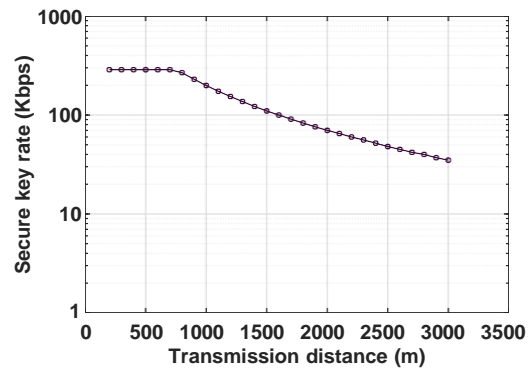

Figure 11: Key rate analysis of QKD framework with existing Alice and Bob set-up.

# 7 Quantum secured video-calling

To showcase a real-life application of a practical QKD system, an end-to-end video calling app, QuViC (Quantum video calling app) has also been developed and integrated with the QKD framework. Using the webcam attached to the Alice and Bob PC terminals and the shared quantum keys available at the Alice's and Bob's terminals for encryption and decryption, this app enables users to place two-way client-to-client quantum secured video call, while executing the QKD protocol in the background.

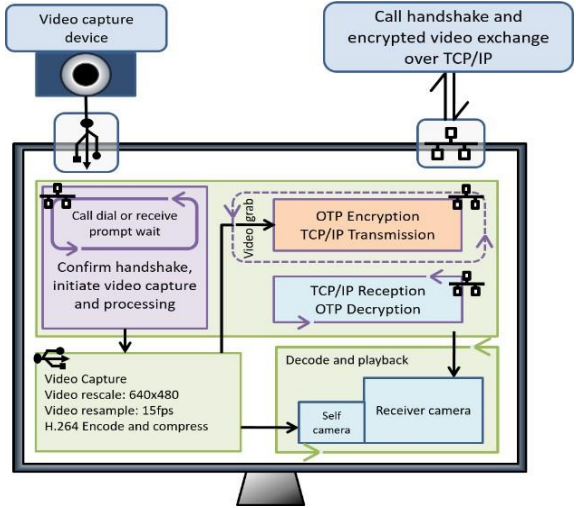

Figure 12: Illustration of QuViC software procedures for placing two-way quantum secured video call.

The end-to-end quantum key encrypted video calling is carried out over TCP/IP link after establishing proper hand-shaking by each party over another TCP link. The video call process can be divided into two sub-processes viz. video transmission and video playback. Transmission sub-process involves video capture (resampling, rescaling and encoding), data fetching for encryption, OTP encryption, TCP/IP transmission. Similarly, video playback sub-process involves TCP/IP reception, OTP decryption, video decode and playback. The hand-shaking for initiating the call happens by a simple parity transmission and confirmation over TCP/IP.

The block schematic illustrating the QuViC software procedures is shown in Fig 12. Using the quantum key generated and distributed through QKD framework, the QuViC app was used to place a video call between the teams sitting at two separate buildings with Alice and Bob's terminals. The call lasted for about 2 mins. and ∼10Mb of quantum key data was consumed in the process, including the fresh quantum keys being generated in the background.

## 8  Conclusion

In this work, we presented fully automated end-to-end QKD framework employing NavIC based synchronizations mechanism. The performance has been validated by demonstrating QKD using BB84 protocol over a distance of 300m of atmospheric channel. This inter-building QKD system demonstrated a low QBER of < 3% with secure key generation rate as high as ∼300 Kbps at an average photon no. per pulse($\mu$) of 0.15. The quantum communication link is operated continuously for several hours during multiple nights. This experimental demonstration has overcome several technological challenges. This manuscript also presented the development of a novel video calling app, QuViC, for two-way client-to-client communication which uses the quantum key generated at both the terminals and demonstrated quantum secured video calling. The entire framework has been designed in such a way that it can be upgraded to work with other GNSS receivers like GPS, GLONASS etc. and also with advanced protocols like decoy state BB84, B92, BBM92 etc.

In future, enhancement in secure key generation rate by increasing the pulse repetition rate will be explored. Video calling app is found to be working well with a lag of few seconds, which can be reduced with further optimization. The future work also includes demonstrating QKD protocols in moving platform based scenarios, emulating satellite based QKD schemes. The decoy based BB84 protocol, which is more robust against eavesdropping attack and capable of achieving higher secure key rate, is also being investigated. All these activities will culminate into the final goal of achieving satellite based quantum communication (SBQC) and distributing secure encryption keys between two Indian ground stations.

## Acknowledgement

The authors would like to acknowledge Shri. K S Parikh, Deputy Director, SNPA and Shri T V S Ram, Group Director, ODCG, for their encouragement and support to this work. We would like to express our gratitude to Shri. R K Bahl, Head, OCD, for regular reviews, timely guidance and many fruitful discussions. Authors are grateful to Smt. Arti Sarkar, Group Director, EOSDIG, Dr. B N Sharma, Head, SSD, Shri. Hriday N Patel, EOPID and Shri Srimanta Mitra, SSD for facilitating mounting and alignment of Bob terminal and support during testing. Authors would also like to acknowledge Shri Nitin Kumar, CDFD, Shri Rakesh K Bijarniya, NRD and Shri Santosh Lacheta, SCTD for extending their support to this work.